\documentclass[aps,prl,reprint,groupedaddress]{revtex4-2}
\usepackage[T1]{fontenc}
\usepackage{graphicx}

\usepackage{amsmath}
\usepackage{bm}
\usepackage{color}

\begin{document}


\title{Metachronal coordination at the mesoscale}


\author{Sebastian Ziegler$^{1,\dagger}$, Megan Delens$^{2,\dagger}$, Ylona Collard$^2$, Maxime Hubert$^{1,3}$, Nicolas Vandewalle$^{2,*}$, and Ana-Sun\v{c}ana Smith$^{1,3,}$}
\email[Corresponding authors: ]{smith@physik.fau.de, nvandewalle@uliege.be}
\email{\newline$^\dagger$ SZ and MD contributed equally}

\affiliation{$^1$PULS Group, Institute for Theoretical Physics, Interdisciplinary Center for Nanostructured Films (IZNF), Friedrich-Alexander-Universität Erlangen-Nürnberg Cauerstr. 3, 91058 Erlangen, Germany}
\affiliation{$^2$GRASP, Institute of Physics B5a, Université de Liège, 4000 Liège, Belgium}
\affiliation{$^3$Group for Computational Life Sciences, Division of Physical Chemistry, Ruđer Bo\v{s}kovi\'{c} Institute, Bijeni\v{c}ka cesta 54, 10000 Zagreb, Croatia}


\begin{abstract}
In nature, metachronal coordination is an efficient strategy for fluid pumping and self-propulsion. Yet, mimetic systems for this type of organization are scarce. Recently, metachronal motion was observed in a bead-based magnetocapillary mesoswimmer, but the mechanism of such device’s behavior remained unclear. Here, we combine theory and experiments to explain two swimming regimes that we identify by modulation of the driving frequency. In the linear, low-frequency regime, the swimmer motion originates from individual bead rotations. However, the high-frequency, metachronal regime is dominated by deformations of the device near a mechanical resonance, which highlights the role of bead inertia for optimized self-propulsion. \end{abstract}

\maketitle

Metachronal coordination is when the moving parts of an organism, like limbs or appendages, move in a sequence with a slight delay between each, creating a wave-like pattern \cite{Byron2021}.
Such behavior is observed for the cilia covering a variety of microorganisms, including Paramecium \cite{Machemer1972, Narematsu2015} and Volvox alga colonies \cite{Brumley2012}.  Metachronal waves were also found to be exhibited by filaments covering the interior of the mammal lung \cite{Yeates1993, Mesdjian2022} or in the extremities of crustaceans or insects \cite{Alben2010, Wilson1965}. The underlying evolutionary principle promoting  the metachronal beat is likely its significantly larger fluid pumping velocity and the energetic efficiency compared to a synchronized beat \cite{Osterman2011, Elgeti2013}. 
Understanding this principle, and establishing a minimal system displaying metachronal waves has therefore attracted considerable attention in the last decade \cite{Palagi2013, Tsumori2016, Hanasoge2018, Dong2020, Gu2020, Collard2020}. 

\begin{figure}
\includegraphics[scale=0.9]{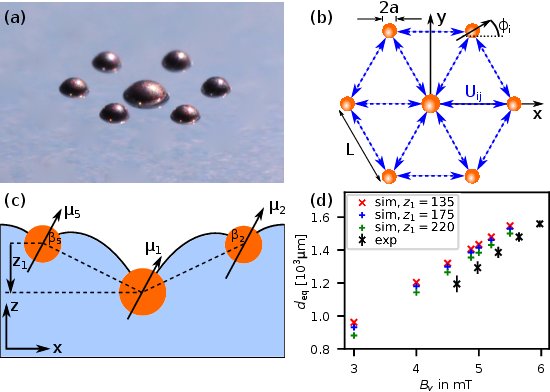}
\caption{The magneto-capillary seven-bead swimmer. (a) Photography of the experimental swimmer. (b) Sketch of the model for the seven-bead swimmer. (c) Cut through the swimmer along the x-direction illustrating the vertical displacement of the central bead and the left-right asymmetric dipole-dipole interactions. (d) Equilibrium distance between two neighboring beads in the seven-bead swimmer in the simulations.}
\label{fig_1}
\end{figure}

Building on our expertise in magnetocapillary swimmers \cite{GrosjeanLagubeauDarras2015, GrosjeanHubertVandewalle2018, GrosjeanHubertLagubeau2016, Grosjean2019}, we have recently shown that it is possible to reconstitute the metachronal organization by driving ferromagnetic beads deposited on the water-air interface with magnetic fields \cite{Collard2020}. However, we did not yet elucidate the mechanism that yields such dynamics, and therefore the optimization of the device remains an open problem. An important hint about the underlying physics is provided by the theoretical generalization of bead assemblies \cite{NajafiGolestanian2004, Felderhof2006, PooleyAlexanderYeomans2007, PandeSmith2015, RizviFarutinMisbah2018, Klotsa2015, Dombrowski2019}. Namely, by quantitative comparison of analytic approaches and experiments \cite{Grosjean2019, SukhovZieglerXie2019, Hubert2021, Sukhov2021}, we showed that magnetocapillary assemblies belong to the class of mesoscale swimmers, where the bead inertia has a considerable impact on its motion, while the fluid inertia remains subordinate \cite{ Gonzalez-Rodriguez2009, Hubert2021, Geyer2022a}. We, therefore, hypothesize that precisely bead inertia plays an important role in metachronal coordination for our magnetocapillary system. The verification of this idea is presented in the current paper.

\emph{System design and equilibrium configuration} Our device (Fig. \ref{fig_1}) consists of a central bead of $800 \ \mathrm{\mu m}$ in diameter, pinned at a water-air interface thanks to capillary effects. It is surrounded by six beads of $500 \ \mathrm{\mu m}$ in diameter. All beads are made of chrome steel alloy AISI 52100 at density $\rho_b = 7.8 \ 10^3 \ \mathrm{kg}/\mathrm{m}^3$. Although the alloy is considered to be a soft ferromagnetic, the particles exhibit a small remanent magnetization of the order of $100 \ \mathrm{A} \, \mathrm{m}^{-1}$ \cite{LagubeauGrosjeanDarras2016}. Therefore, the total moment of bead $i$ is $\bm{\mu}_i = \bm{\mu}_{rem, i} + 4 \pi a_i^3 \chi \bm{B}/(3 \mu_0)$, with $a_i$ and $\chi = 3$ being the bead's radius and susceptibility, $\mu_0$ the vacuum permeability, and $\bm{\mu}_{rem, i}$ the remanent moment. 
The magnetic induction field $\bm{B}$ is applied on the device by three pairs of orthogonal Helmholtz coils, each providing up to $8 \ \mathrm{mT}$. The currents in the $x$- and $y$-coils are produced by an arbitrary function generator and a pair of amplifiers, to eventually propel the swimmer. The vertical field is typically kept constant at $B_v=4.9 \ \mathrm{mT}$. The imposed $\bm{B}$ field, therefore, yields a magnetic potential $U^m$ between the beads
\begin{eqnarray}
&&U^{m} = - \sum_i \bm{\mu}_i \cdot \bm{B} \label{eq:magneticPotential}\\
&&+ \frac{\mu_0}{8\pi} \sum_{i} \sum_{j \neq i}  \frac{\bm{\mu}_i \cdot \bm{\mu}_j - 3 (\bm{\mu}_i \cdot \bm{e}_{ij}) (\bm{\mu}_j \cdot \bm{e}_{ij})}{|\bm{r}_{ij}|^3}, \nonumber 
\end{eqnarray}
with $\bm{e}_{ij}$ the unit vector along $\bm{r}_{ij} := \bm{r}_j - \bm{r}_i$, and $i, j$ bead indices.  Capillary effects bind the beads strongly at the contact line with the liquid interface \cite{Grosjean2019}. As a result, the beads mostly move in the $xy$-plane and rotate around the $z$-axis (with a relative orientation to the $x$-axis described by $\phi_i$, see Fig. \ref{fig_1}(b)), while experiencing repulsive magnetic interactions.  

Because the beads deform the water-air interface, they also experience an attractive capillary potential $U^c$ \cite{Kralchevsky2000}, which takes the form 
\begin{equation}
U^\mathrm{c} = -\pi \sigma \sum_i \sum_{j \neq i} q_i q_j K_0 \left(|\bm{r}_{ij}^\parallel|/l_c \right). 
\label{eq:capillaryPotential}
\end{equation}
Here $\sigma = 72 \ \mathrm{mN/m}$ is the water surface tension, $K_0$ is the modified Bessel function of the second kind, and $\bm{r}_{ij}^\parallel$ is the projection of $\bm{r}_{ij}$ on the $xy$-plane. Moreover, with $g$ being the gravitational constant, and $\rho$ the fluid density, $l_c = \sqrt{\sigma / g \rho}$ is the capillary length, which, for the water-air interface, is $l_c\approx2.7$ mm. Finally, the capillary charge $q_i$ represents the characteristic length of the vertical deformation of the liquid interface around each particle, which for the current case is around $12 \ \mathrm{\mu m}$ for $500 \ \mathrm{\mu m}$ beads and $45 \ \mathrm{\mu m}$ for $800 \ \mathrm{\mu m}$ beads \cite{Collard2022a}. This difference in $q_i$ contributes to the overall deformation of the interface as comprised in $z_1$ (Fig. \ref{fig_1}(c)), which will lead to asymmetric swimmer deformation in a horizontal field (Fig. \ref{fig_1}(c)), inducing locomotion.

Application of only the vertical field relaxes the seven-bead system to its equilibrium configuration with a six-fold symmetry, as shown in Fig. \ref{fig_1}(a). The equilibrium configuration, characterized by the center-center distance between the outer beads and the central bead, is measured to be $1293 \ \mathrm{\mu m}$ for $B_v = 4.98 \ \mathrm{mT}$ and $1386 \ \mathrm{\mu m}$ for $B_v = 5.31 \ \mathrm{mT}$. Similar values are obtained theoretically, with the approximation that capillary interactions depend only on the projection $\bm{r}_{ij}^{\parallel}$ onto the liquid interface (Fig. \ref{fig_1}(d)). Under these circumstances, we find that the equilibrium distance increases quasi-linearly with $B_v$ as a result of the increasing magnetic repulsion (Fig. \ref{fig_1}(d)).

\emph{A theoretical model for magnetocapillary devices} In order to model the behavior of our mesoswimmer, we consider the hydrodynamic interactions between the beads using the bulk Rotne-Prager tensor, as well as magnetic and capillary interactions as shown in Eqs. \eqref{eq:magneticPotential} and \eqref{eq:capillaryPotential}. Following the strategy for mesoswimmers presented in our earlier work \cite{Hubert2021}, we also include the beads' masses $m_i = 4/3 a_i^3 \rho_b \pi$ and moments of inertia $I_i = 2/5 a_i^2 m_i$. The ensuing equations of motion are given by 
\begin{equation}
\begin{pmatrix}
m_i \ddot{\bm{r}}_i \\
I_i \ddot{\phi_i}
\end{pmatrix}
 = \sum_{j = 1}^7
\begin{pmatrix}
\hat{R}_{ij}^{tt} & \hat{R}_{ij}^{tr} \\
\hat{R}_{ij}^{rt} & \hat{R}_{ij}^{rr}
\end{pmatrix}
\cdot 
\begin{pmatrix}
\dot{\bm{r}_j} \\
\dot{\phi_j}
\end{pmatrix} 
 + \begin{pmatrix}
\bm{F}^{c}_{i} \\
0
\end{pmatrix} +
\begin{pmatrix}
\bm{F}^{m}_{i} \\
T^{m}_{i}
\end{pmatrix},
\label{eq:EOM}
\end{equation}
with $\hat{R}$ the resistance matrix (see SI Section I for the details), $\bm{F}^{c}_{i}$ the capillary forces, and $\bm{F}^{m}_{i}$ and $T^{m}_{i}$ magnetic forces and torques. This equation is solved numerically using an algorithm with adaptive step size as implemented in \textit{Mathematica} \cite{Mathematica2020}. We perform the full eigenmode analysis where we find 11 deformation modes with eigenfrequencies between $1 \ \mathrm{Hz}$ and $5.5 \ \mathrm{Hz}$ (SI Section II). 

\begin{figure*}
\includegraphics[scale=1]{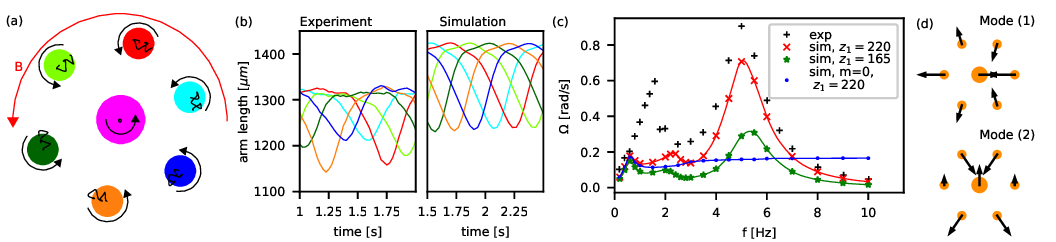}
\caption{Seven-bead swimmer in a rotating magnetic field. (a) Numerical bead trajectories measured over two periods of the field, the bead rotation is depicted by black arrows. (b) Time-dependent arm lengths of the swimmer in the experiments (left) and in the simulations (right). The color corresponds to the bead color in (a). (c) Frequency-dependent angular velocity of the swimmer in the experiments (black pluses) and in the simulations for different values of the depth of the central bead $z_1$ (red crosses and green starts) as well as for zero mass (blue points). The extrapolated lines are a guide for the eye. (d) The two most excited degenerate eigenmodes in the case of the rotating field.}
\label{fig_2}
\end{figure*}

\emph{Device in the rotational field} To reconstitute the metachronal motion we start with a simple rotating magnetic field of the form 
\begin{equation}
B_x = B_h \cos(2 \pi f t), \ B_y = B_h \sin(2 \pi f t), 
\label{eq:rotatingField}
\end{equation}
with $f$ the frequency and $B_h = 0.742 \ \mathrm{mT}$ the magnitude of the horizontal field. 
As a result, we observe metachronal arm compressions and extensions, traveling in a rotating fashion around the swimmer. Thereby, the outer beads move both inward and toward the next outer bead in the direction of rotation of the magnetic field (Fig. \ref{fig_2}(a) and \ref{fig_2}(b)). Notably, the vertical depth $z_1$ of the central bead breaks the axis-symmetry of the dipole-dipole interactions, giving rise to only one metachronal wave instead of two (Fig. \ref{fig_1}(c)). Furthermore, our calculations reveal that each bead rotates around the $z$-axis as a consequence of their remanent magnetic moment coupling to the external $\bm{B}$-field (Fig. \ref{fig_2}(a)). Actually, these bead rotations transform the overall torque applied by the field into the overall mesoswimmer rotation via hydrodynamic interactions, while bead displacements play little role. The obtained angular velocity of the mesoswimmer $\Omega$ is in the direction of the $\bm{B}$-field and increases for larger $z_1$ (compare green stars and red x symbols in Fig. \ref{fig_2}). Translations are not present as there is no net force and because the device is rotationally symmetric. 

We now explore in experiments and in calculations the way in which the angular velocity of the mesoswimmer $\Omega$ depends on the driving frequency $f$. We systematically observe a low- and a high-frequency regime. At low frequencies, we find a linear increase of $\Omega$ towards a peak, which in experiments appears at $f \approx 1 - 2 \ \mathrm{Hz}$ (black crosses in Fig. \ref{fig_2}). The calculations reveal that at low frequencies, the rotations of the individual beads instantaneously follow the field, producing hydrodynamic interactions which induce the device rotation. The higher the field, the stronger the interactions and the faster the swimmer rotation. As the frequency increases further, the bead rotations can no longer follow the field, and $\Omega$ drops.  We, therefore, find that the position of the low-frequency peak can be associated with the step-out frequency \cite{Mahoney2014}
\begin{equation}
f_{so, i} = \frac{1}{2 \pi} \frac{|\bm{\mu}_{rem, i}| B_h}{8 \pi \eta a_i^3},
\end{equation}
which, for the theoretical model, is calculated as $f_{so} \approx 0.7 \ \mathrm{Hz}$ for both bead sizes. The higher step-out frequency observed experimentally is a result of the decreased hydrodynamic friction due to partial bead immersion in the fluid. Indeed by theoretically scaling down the hydrodynamic friction (first term in Eq. \eqref{eq:EOM}), the linear regime is extended also numerically (SI Fig. 2). Furthermore, to check if the bead mass plays a role in this regime, we set $m_i=0$ in Eq. \eqref{eq:EOM} and find no effect of inertia (blue dots in Fig. \ref{fig_2}).   

At high frequencies, we experimentally find another peak at around $5 \ \mathrm{Hz}$ (Fig. \ref{fig_2}(c)). Notably, this peak is absent when setting the bead masses to zero in the simulations (Fig. \ref{fig_2}(c)), but is present even when the remanent magnetic moments are set to zero (SI Fig. 3). This suggests that this peak is the consequence of the translations of the beads in the $xy$-plane. We, therefore, revert to the natural frequency analysis of the device (SI Section II). We find that two main deformation modes, associated with the eigenfrequency of $5.5 \ \mathrm{Hz}$, are excited near the high-frequency peak (SI Fig. 4). This is in excellent agreement with the maximal swimming speed observed in both experiments and calculations. We conclude that this regime is indeed associated with a mechanical inertia-induced resonance of the magnetocapillary system.

\begin{figure*}
\includegraphics[scale=1]{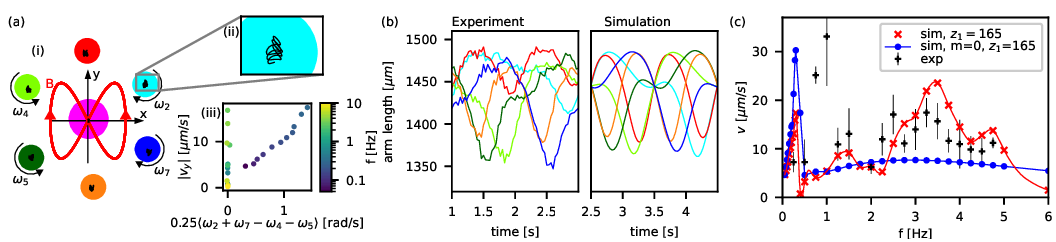}
\caption{Seven-bead swimmer in a Lissajous-type magnetic field. (a) Simulated bead trajectories for three periods of the field ((i) and (ii)). Inset (iii) shows the dependence of swimming speed in $y$-direction on the average angular velocities of the four side beads. At low frequencies, a linear relation is observed suggesting that swimming speed and bead rotation are proportional. (b) Time-dependent arm lengths of the swimmer in the experiments (left) and in the simulations (right) for $f = 1.0 \ \mathrm{Hz}, B_h = 0.5 \ \mathrm{mT}$, colors are identical to (a). (c) Comparison of the frequency-dependent swimming speed of the swimmer in the experiments and simulations for $B_h = 1.2 \ \mathrm{mT}$.}
\label{fig_3}
\end{figure*}

\emph{Efficient metachronal device} While the simple rotational field was able to provide a minimal model for metachronal organization, very low pumping, and no swimming were achieved. To improve on this point, we apply a Lissajous-type field 
\begin{equation}
B_x = B_h \cos(2 \pi f t), \ B_y = B_h \sin(4 \pi f t),
\label{eq:LissajousField}
\end{equation}
with $B_h = 1.2 \ \mathrm{mT}$. 
Doubling the frequency along the $y$-axis leads to very efficient translational locomotion at speeds up to tens of $\mathrm{\mu m/s}$ along the $y$-direction, in a force-free manner. This driving also produces a vanishing average torque, hence no net rotations of the device are observed. We again recover the low- and high-frequencies response (Fig. \ref{fig_3}(c)). In the low-frequency, linear regime (up to 1 Hz), the self-propulsion is again a result of bead rotations. The strongest contribution comes from the four beads on the sides of the swimmer oriented in the direction of motion along the $y$-axis (green and blue beads with indicated rotations in Fig. \ref{fig_3}(a)). These four beads rotate continuously in the same direction because of the alternating action of (i) the magnetic torques induced by the external field, and (ii) torques due to dipole-dipole interactions (SI Section IV and SI Fig. 5). The induced rotlets add up to a flow in the positive $y$-direction resulting in the self-propulsion of the swimmer. Given the linear response of beads, the strength of the rotlet is proportional to frequency, and consequently, an increase in the swimming speed is observed (Fig. \ref{fig_3}(c) and SI Fig. 6). Despite not being related to swimmer deformations, the latter reaches up to $35 \ \mathrm{\mu m/s}$, before the beads stop responding instantaneously to the field changes.

 In the high-frequency regime ($3 - 5 \ \mathrm{Hz}$), the motion of the device is a direct consequence of a metachronal organization of the beads' motions. With Lissajous driving, the compressions of the swimmer side arms (blue and green beads) travel from the back to the front of the device alternately on both sides of the swimmer with the frequency $f$, i.e. the frequency of $B_x$ (Eq. \eqref{eq:LissajousField} and Fig. \ref{fig_3}(a) and (b)). In contrast, the distances of the front (red) and the back (orange) beads to the central one oscillate with $2f$, the frequency of $B_y$. Like with the rotational field, the first two eigenmodes contribute the most. However, due to stronger driving fields, the excitation of two additional degenerate mode pairs is observed (see SI Fig. 1 for modes 4 and 5, as well as 10 and 11, and SI Fig. 7 for the decomposition). 
The peak in the high-frequency regime is recovered in simulations with zero remanent moments (SI Fig. 6), i.e. at no bead rotations. This demonstrates that those modes are indeed responsible for the locomotion of the device in the high-frequency regime.

The observed peaks are, notably, at lower frequencies than with rotational $\bm{B}$-fields, because of the frequency doubling of the magnetic field in $y$-direction. Furthermore, given the considerable speed of the device, it is instructive to verify the characteristic Reynolds numbers \cite{Hubert2021, Derr2022}. For the fluid, with $L$ being the total length of the swimmer, and $\eta$ the fluid viscosity, $Re_f =\rho L v/ \eta\approx 0.09$, which is well within the low $Re$ regime. On the other hand, the swimmer's Reynolds number is $Re_s = 2 \pi \rho_b a_i^2 f/\eta\approx 10 - 30$ near the resonance. This highlights the mesoscale nature of the swimmer. 
To illustrate this result, we show that the high-frequency peak is lost in calculations with zero bead masses (Fig. \ref{fig_3}(c)), while metachronal coordination, although at lower amplitude, persists. Consequently, we conclude that the beads' inertia can amplify the metachronal stroke. The consequences are enhanced rotational and translational locomotion.


\emph{Conclusions} In this paper we clearly identified the mechanisms by which magnetic, capillary, and hydrodynamic forces acting on seven beads on an interface induce self-propulsion. Our bead assembly generally shows two regimes of locomotion depending on the driving frequency. At low frequencies, the linear coupling of the $\bm{B}$-field to the remanent magnetic moments of the beads induces rotations where beads perfectly follow the oscillating field. Depending on the driving protocol, this may give rise to both efficient rotation and translation. 
By type, this mechanism falls into the class of theoretically suggested microswimmers with components that act on passive parts of the assembly by creating flow fields that ultimately drive the device \cite{HoellLoewenMenzel2017, Hoell2018}. 
Our results, however, show that such a mechanism can also be exploited in an experimental setting and used to efficiently generate translations and rotations of the device.

The metachronal organization represented by oscillatory consecutive translations of beads is, however, reconstituted at high driving frequencies. By quantitative comparison of experiments and theory, we show that these bead motions strongly depend on the mesoscale nature of the swimmer. Specifically, the inertia of the beads is found to enhance the swimming stroke while the natural mechanical resonances of the swimmer are excited. This clearly demonstrates that exploiting high-density constituents in yet small-scale swimmers presents a pathway to circumvent the often observed slow locomotion of massless swimmers. Our results may also be relevant to understand the common occurrence of metachronal coordination in a number of mesoscale organisms such as crustaceans. A nice example comes from krill, which exploit metachronal coordination, while clearly being in the mesoswimming regime \cite{Zhang2014}.
Additional investigation is, however, necessary to explore the relation between metachronal organization and inertial effects on a broader variety of living species. Furthermore, both swimming mechanisms unveiled in this work might prove useful in the future design of artificial microswimmers or microscopic pumps in technological applications. 

\begin{acknowledgements}
This work is financially supported by the FNRS CDR project number J.0186.23 entitled ``Magnetocapillary Interactions for Locomotion at Liquid Interfaces" (MILLI). 
\end{acknowledgements}

\bibliography{main}

\end{document}



\title{Supplementary information for the article \\
``Metachronal coordination at the mesoscale''
}


\author{Sebastian Ziegler$^{1,\dagger}$, Megan Delens$^{2,\dagger}$, Ylona Collard$^2$, Maxime Hubert$^{1,3}$, Nicolas Vandewalle$^{2,*}$, and Ana-Sun\v{c}ana Smith$^{1,3,}$}
\email[Corresponding author:]{smith@physik.fau.de, nvandewalle@uliege.be}
\thanks{\newline$ ^\dagger$SZ and MD contributed equally}

\affiliation{$^1$PULS Group, Institute for Theoretical Physics, Interdisciplinary Center for Nanostructured Films (IZNF), Friedrich-Alexander-Universität Erlangen-Nürnberg Cauerstr. 3, 91058 Erlangen, Germany}
\affiliation{$^2$GRASP, Institute of Physics B5a, Université de Liège, 4000 Liège, Belgium}
\affiliation{$^3$Group for Computational Life Sciences, Division of Physical Chemistry, Ruđer Bo\v{s}kovi\'{c} Institute, Bijeni\v{c}ka cesta 54, 10000 Zagreb, Croatia}


\begin{abstract}
\end{abstract}


\maketitle

\section{Details on the numerical calculations}\label{section_details_sim}
In the simulations, the bulk low Reynolds number hydrodynamics at the level of the Rotne-Prager approximation, i.e. including all terms up to the order $\mathcal{O}\left(a/r\right)^3$, is used. Since at the Rotne-Prager level, the expressions for the mobility matrix $\hat{\mu}$ are simpler than for the resistance matrix $\hat{R} = - \hat{\mu}^{-1}$ \cite{KimKarilla1991}, we implement in the simulations the mobility matrix and invert it later. 
The corresponding expressions for components of the mobility matrix are \cite{Dhont1996, PicklPandeKoestler2017}: 
\begin{eqnarray}
&&\hat{\mu}^{tt}_{ij} = \frac{1}{8 \pi \eta |\bm{r}_{ij}|}  \left(\hat{1} + \frac{\bm{r}_{ij} \otimes \bm{r}_{ij}}{|\bm{r}_{ij}|^2} \right) + \nonumber \\
&&\hspace{2 cm}\frac{a_i^2 + a_j^2}{24 \pi \eta |\bm{r}_{ij}|^3} \left(\hat{1} - 3 \frac{\bm{r}_{ij} \otimes \bm{r}_{ij}}{|\bm{r}_{ij}|^2} \right) \ (i \neq j), \nonumber	\\
&&\hat{\mu}^{tt}_{ii} = \frac{\hat{1}}{6 \pi \eta a_i}, \\
&&\hat{\mu}^{tr}_{ij} = \frac{1}{8 \pi \eta |\bm{r}_{ij}|^3} \left(\bm{r}_{ij} \times \right)  \ (i \neq j), \ \hat{\mu}^{tr}_{ii} = \hat{0}, \nonumber 
\hat{\mu}^{rt}_{ij} = \left( \hat{\mu}^{tr}_{ji} \right)^T, \\
&&\hat{\mu}^{rr}_{ij} = 0  \ (i \neq j), \ \hat{\mu}^{rr}_{ii} = \frac{1}{8 \pi \eta a_i^3}. \nonumber 
\end{eqnarray}
Here, upper indices $t$ and $r$ denote translation and rotation, $i$ and $j$ refer to two swimmer beads. The symbol $\otimes$ denotes the tensor product of two vectors and $\left(\bm{r}_{ij} \times \right)$ is a rank-two tensor defined as 
\begin{equation}
\left( \bm{r}_{ij} \times \right):= \epsilon_{pqr} \left( \bm{r}_{ij} \right)_q \bm{e}_p \bm{e}_r,
\end{equation}
with $\epsilon$ the Levi-Civita symbol, $p, q, r$ spatial indices, and $\bm{e}_p$ the unit vector in the direction associated with the index $p$. The water viscosity $\eta$ is assumed as $\eta = 1.0 \ \mathrm{mPa \, s}$. 

Since we consider rotations only around the $z$-axis, we use for the components of the mobility matrix of the form $\hat{\mu}^{tr}_{ij}$, $\hat{\mu}^{rt}_{ij}$ and $\hat{\mu}^{rr}_{ij}$ only the entries associated with rotation about the $z$-axis. For instance, $\hat{\mu}^{rr}_{ij}$ then becomes a scalar while $\hat{\mu}^{tr}_{ij}$ and $\hat{\mu}^{rt}_{ij}$ are represented as tensors of dimensionality $3 \times 1$ and $1 \times 3$, respectively. 

We then define the full mobility matrix as 
\begin{equation}
\hat{\mu} = 
\begin{pmatrix}
\hat{\mu}^{tt}_{11} & \ldots & \hat{\mu}^{tt}_{17} & \hat{\mu}^{tr}_{11} & \ldots & \hat{\mu}^{tr}_{17} \\
\vdots & \ddots & \vdots & \vdots &  \ddots & \vdots \\
\hat{\mu}^{tt}_{71} & \ldots & \hat{\mu}^{tt}_{77} & \hat{\mu}^{tr}_{71} & \ldots & \hat{\mu}^{tr}_{77} \\
\hat{\mu}^{rt}_{11} & \ldots & \hat{\mu}^{rt}_{17} & \hat{\mu}^{rr}_{11} & \ldots & \hat{\mu}^{rr}_{17} \\
\vdots &  \ddots & \vdots & \vdots &  \ddots & \vdots \\
\hat{\mu}^{rt}_{71} & \ldots & \hat{\mu}^{rt}_{77} & \hat{\mu}^{rr}_{71} & \ldots & \hat{\mu}^{rr}_{77} \\
\end{pmatrix}.
\end{equation}
Inverting this expression gives rise to the components of the resistance matrix $\hat{R}$ that are used in Eq. (3) of the main text. 

The magnetic forces derive from the potential given in Eq. (1) in the main text. As a result, the magnetic forces $\bm{F}_i^m$ generally would also have a component in $z$-direction. However, under the assumption that the capillary forces bind the particles at their contact line with the liquid interface, we neglect the $z$-component of the magnetic forces in our simulations.  
The remanent magnetic moments are set in the simulations to $\mu_\mathrm{rem} = 9.35 \cdot 10^{-9} \ \mathrm{A \, m^2}$ for the bead of diameter $800 \ \mathrm{\mu m}$ and to $\mu_\mathrm{rem} = 2.34 \cdot 10^{-9} \ \mathrm{A \, m^2}$ for the beads of diameter $500 \ \mathrm{\mu m}$ \cite{Hubert2018}. Importantly, these experimental values represent the in-plane components of the remanent magnetic moments. Consistently, we impose in the simulations that the remanent moment is parallel to the interface plane by choosing in-plane initial conditions and the assumption that the beads can only rotate in-plane.
The capillary forces, as defined from the potential in Eq. (2) of the main text, act only within the $xy$-plane. 

All beads are initialized with zero velocities in a six-fold symmetric structure with the large central bead in the origin. 
The initial remanent moments of the beads are oriented parallel to the initial horizontal magnetic field. 
The equation of motion for magnetocapillary swimmers is given by Eq. (3) in the main text. This equation of motion is then solved numerically using Mathematica's \cite{Mathematica2020} \textit{NDSolve} method, which is using an algorithm with adaptive step size. 

\section{Eigenmode decomposition of the swimmer deformation}
\begin{figure*}
\includegraphics[scale=1]{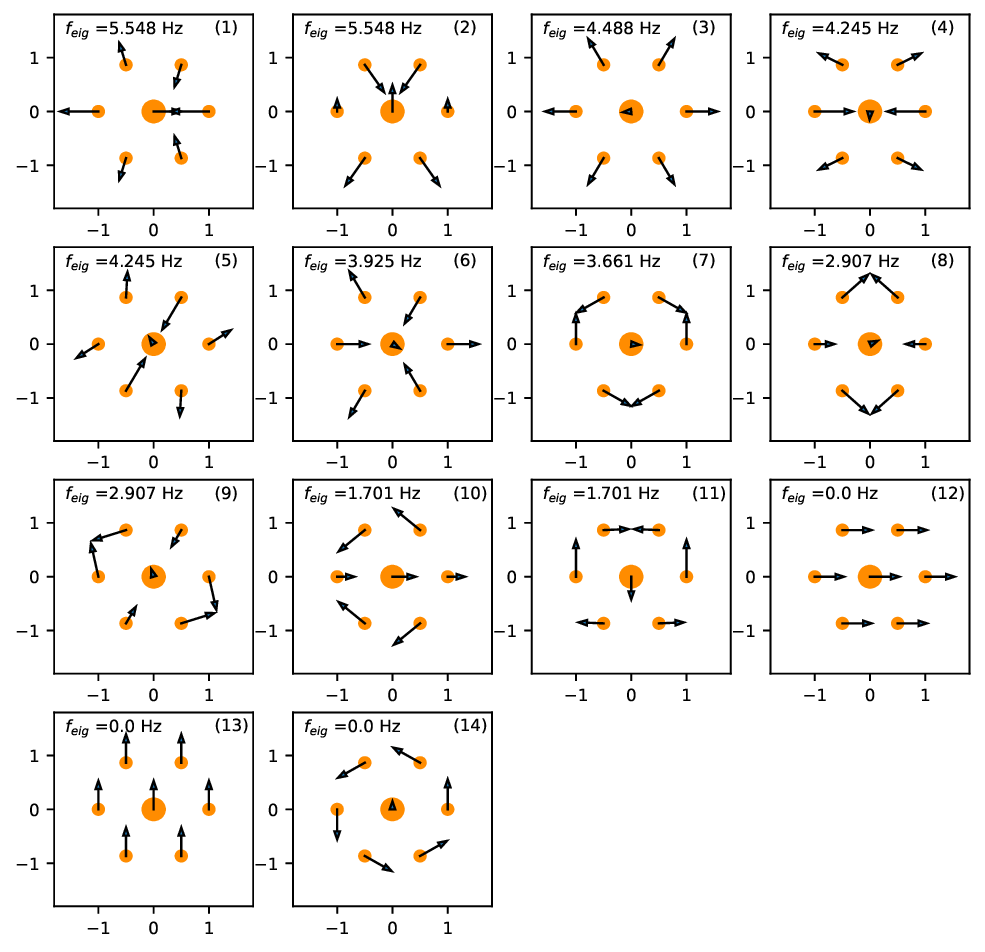}
\caption{Sketch of the eigenmodes of the magnetocapillary swimmer with the respective eigenfrequencies of the modes.}
\label{fig_eigenmodes}
\end{figure*}
To determine the eigenmodes of the magnetocapillary seven-bead swimmer, we consider the swimmer in a vertical magnetic field of $B_v = 4.9 \ \mathrm{mT}$. After the swimmer has been simulated for $10 \ s$ to reach the equilibrium configuration, the linear magnetocapillary forces associated with an infinitesimal bead displacement in the $xy$-plane are calculated, returning a $14 \times 14$ matrix $\underline{M}$, to be contracted with the bead displacements. 
To obtain the associated bead accelerations, we calculate
\begin{equation}
\underline{K} := 
\begin{pmatrix}
m_1 & 0 & \ldots & 0 & 0\\
0 & m_1 & 0 & \ldots & 0\\
\vdots & \vdots & \ddots & \vdots & \vdots \\
0 & \ldots & 0 & m_7 & 0\\
0 & \ldots & 0 & 0 & m_7\\
\end{pmatrix}
\cdot \underline{M}, 
\end{equation}
the matrix relating a bead displacement with the corresponding restoring accelerations experienced by the beads. 
The eigenvectors of the matrix $\underline{K}$ are then the eigenmodes of the system, while the associated eigenvalues $\lambda$ give rise to corresponding eigenfrequencies $f = \sqrt{-\lambda} / (2 \pi)$. 

The resulting eigenmodes and the associated eigenfrequencies are shown in Fig. \ref{fig_eigenmodes}. The first 11 modes correspond to swimmer deformations and thus to non-zero eigenfrequencies, while the remaining three modes correspond to translation and rotation. There exist four pairs of degenerate deformation eigenmodes associated with the same eigenfrequency, of which each pair spans a two-dimensional subspace of degenerate eigenmodes. For each of these degenerate subspaces, the eigenmodes shown in Fig. \ref{fig_eigenmodes} have been chosen as two linearly independent modes of highest possible symmetry. 

For the decomposition of the swimmer deformation in eigenmodes, the displacement of each bead in the swimmer's frame of reference is required for each instant of time. Since the beads cannot move in the $z$-direction by assumption, we restrict our focus to the $x$- and $y$-components of the bead position vectors $\bm{r}_k$. 
We calculate the swimmer orientation for each time step by 
\begin{equation}
\alpha_s = \mathrm{arg}\left(\sum_{k = 2}^7 \exp\left(i (\alpha_k - \frac{2 \pi}{6} (k - 2)\right) \right),
\end{equation}
with $\alpha_k$ the orientation of $\bm{r}_k - \bm{r}_1$, i.e. the orientation of the arm associated with bead $k$, with $k \in \{2, ..., 7\}$, and $\mathrm{arg}$ the argument of a complex number. 
We then calculate the center of mass 
\begin{equation}
\bm{r}_{COM} := \frac{\sum_{k=1}^7 m_k \bm{r}_k}{\sum_{k=1}^7 m_k},
\end{equation}
as well as the rotated bead positions such that the average swimmer orientation is zero and the center of mass is in the origin of the coordinate system,
\begin{equation}
\bm{r}^\prime_k = \hat{R}(-\alpha_s) \cdot (\bm{r}_k - \bm{r}_{COM}).
\end{equation}
Here, 
\begin{equation}
\hat{R}(\alpha) := 
\begin{pmatrix}
\cos(\alpha) & -\sin(\alpha) \\
\sin(\alpha) & \cos(\alpha)
\end{pmatrix},
\end{equation}
denotes the general $2 \times 2$ rotation matrix. 

The bead displacement is then calculated as the difference between each $\bm{r}^\prime_k$ and the symmetric six-fold structure with the central bead in the origin and all other bead positions defined by
\begin{equation}
\bm{r}^{eq}_k = l^{eq} \left(\cos\left(\frac{2 \pi}{6} (k-2) \right), \sin\left(\frac{2 \pi}{6} (k-2) \right)  \right)^T. 
\end{equation} 
Here, $l^{eq}$ denotes the swimmers time-averaged arm length.

\section{The magnetocapillary swimmer in a rotating magnetic field}
\begin{figure}
\includegraphics[scale=1.2]{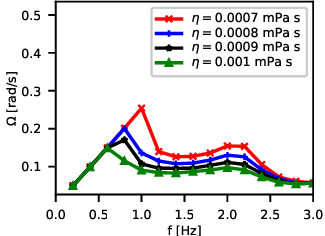}
\caption{Frequency-dependent rotation speed of different values of the viscosity. A decrease in viscosity is accompanied by an increasing step-out frequency $f_{so}$}
\label{fig_Rotation_viscDep}
\end{figure}
The peak in the linearly growing low-frequency regime is determined by the step-out frequency $f_{so}$ (Eq. (5) in the main text). As expected from the analytical expression, the low-frequency regime extends to higher frequencies when the viscosity of the fluid is decreased (Fig. \ref{fig_Rotation_viscDep}). Since in the experiments the beads are only partially immersed in the fluid, the corresponding effective viscosity is lower. As a consequence, the low-frequency regime extends to higher frequencies in the experiments than in our simulations which assume bulk hydrodynamics. 

\begin{figure}
\includegraphics[scale=0.45]{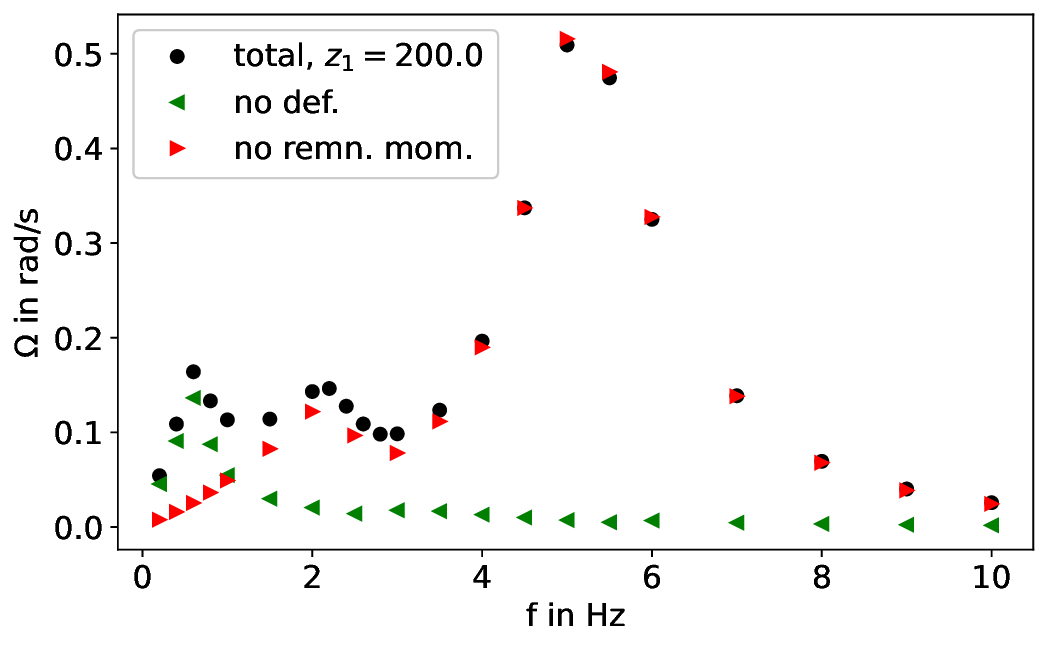}
\caption{Frequency-dependent rotation speed of the magnetocapillary swimmer in the rotating magnetic field in the full simulations (black dots), in the absence of forces deforming the swimmer (green left-pointing triangles) and in the absence of remnant moments (red right-pointing triangles).}
\label{fig_mechanism_rotation}
\end{figure}
To illustrate the origin of swimmer rotation in a rotating magnetic field, we perform simulations of the magnetocapillary system with zero remanent magnetic moments (red right-pointing triangles in Fig. \ref{fig_mechanism_rotation}), as well as without deforming forces (green left-pointing triangles in Fig. \ref{fig_mechanism_rotation}). The latter case was achieved by setting the horizontal magnetic field strength to zero for the particle forces, while keeping the non-zero field strength for the particle torques. 
We observe that the swimmer rotation in the full simulations (black dots in Fig. \ref{fig_mechanism_rotation}) is reproduced in the low frequency regime in the case of no deforming forces, while the first peak is lost in the case of zero remanent moments. For frequencies above $2 \ \mathrm{Hz}$, the main contribution to the swimmer's rotation is recovered in the simulations with absent remanent moments. Thus, in the high-frequency regime, the rotations of single beads are subordinate and the swimmer rotation emerges from the metachronal motion of the beads in the interface plane.

\begin{figure}
\includegraphics[scale=1]{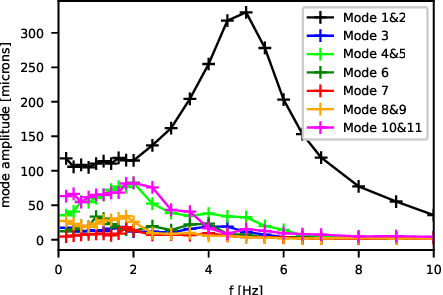}
\caption{Excitation of the eigenmodes of the magnetocapillary swimmer in the rotating magnetic field in the experiments, degenerate eigenmodes associated to the same eigenvalue are combined into a single curve. The lines are a guide for the eye.}
\label{fig_eigModesRot}
\end{figure}
In the high-frequency regime, the swimmer deformations are associated primarily to the first two eigenmodes (see Fig. \ref{fig_eigenmodes}). To show this, we decompose in Fig. \ref{fig_eigModesRot} the excitation of the experimental magneto-capillary swimmer's deformation in the eigenmodes in dependence of the frequency of the rotating magnetic field $f$. While at low $f$, several modes are excited, the degenerate pair of eigenmodes 1 and 2 is dominant in the high-frequency regime and in particular near the mechanical resonance at $f \approx 5 \ \mathrm{Hz}$. The position of the high-frequency peak as well as the position of maximal mode excitation is in good agreement with the eigenfrequency associated to these two modes, which is $f \approx 5.5 \ \mathrm{Hz}$.

\section{The magnetocapillary swimmer in a Lissajous-type magnetic field}
\begin{figure}
\includegraphics[scale=0.55]{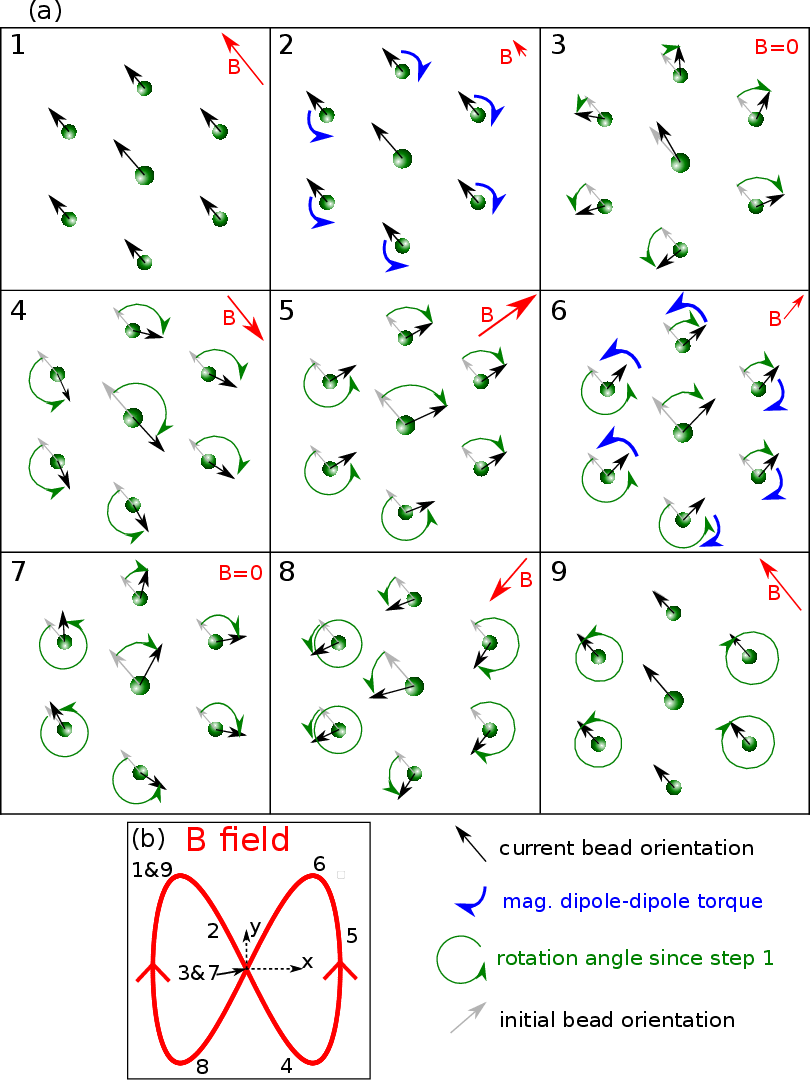}
\caption{Swimming mechanism of the seven-bead swimmer in the low-frequency regime. (a) Sketch of the orientations of the remanent moments during one full cycle of the Lissajous-type horizontal magnetic field. (b) Plot of the Lissajous field, with the $\bm{B}$-field of the different snap shots in part (a) indicated.}
\label{fig_Lissajous_single_bead_rot}
\end{figure}
Our simulations allow to better understand the self-propulsion mechanism at low frequencies, as they enable us to observe the orientation of each bead, which is very difficult to implement in the experiments. 
At low frequencies, we find that the swimmer rotation results from continuous rotations of outer beads around the $z$-axis. In Fig. \ref{fig_Lissajous_single_bead_rot}, we sketch the mechanism that leads to the single bead rotations in the simulations. Here, the black arrows indicate the current bead orientation while the green arrows indicate the cumulative rotation since step 1. The pale gray arrows indicate the initial bead orientation and the blue arrows indicate dipole-dipole torques. 

Within the Lissajous magnetic field, periods during which the beads' orientations follow the magnetic field alternate with periods where the Lissajous field is weak and the bead rotations result from the magnetic dipole-dipole interactions of the beads. While a strong external magnetic field leads to bead rotations in the same direction for all beads, the dipole-dipole interactions at instants of a weak external field are roughly axis-symmetric with respect to the preceding orientation of the external magnetic field. This leads to a breaking of the rotational symmetry and enables swimmer translation. 

In detail, when the horizontal magnetic field switches direction, the three beads on the lower left side rotate counterclockwise (CCW), while the others rotate clockwise (CW) (Fig. \ref{fig_Lissajous_single_bead_rot}, step 2). After the external field as switched direction (step 3), all beads follow the CCW rotation of the external field (step 5). When the horizontal field switches direction again (steps 6 and 7), the orientation of the remanent bead moments is again aligned with the preceding orientation of the magnetic field, which is, however, different from that of the first switching event. As a consequence, the three beads on the upper left side rotate CCW while the beads on the lower left side rotate CW. After the horizontal field has switched direction, all beads again follow the CW rotation of the magnetic field. 
After a full cycle of the Lissajous-type magnetic field, the two beads on the swimmer's left side have performed a full CCW rotation, while the beads on the right side have undergone a full CW rotation. The resulting flow field points upwards at the central bead as well as at the beads above and below it. Consequently, this leads to upwards swimming of the magnetocapillary ensemble.

\begin{figure}
\includegraphics[scale=0.45]{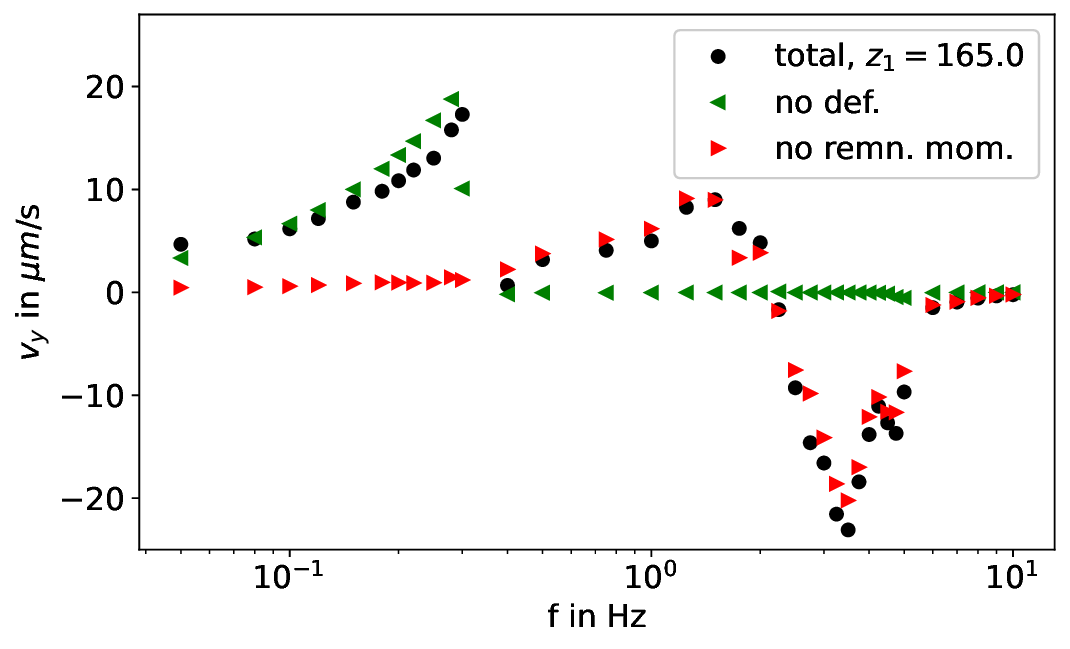}
\caption{Frequency-dependent swimming velocity of the magnetocapillary swimmer in the Lissajous field in full simulations (black dots), in the absence of forces deforming the swimmer (green left-pointing triangles) and in the absence of remnant moments (red right-pointing triangles).}
\label{fig_mechanism_Lissajous}
\end{figure}
Similarly to the case of the rotating magnetic field, the efficient propulsion is observed in both the low-frequency regime detailed above as well as a high-frequency regime, here at $f \approx 3 - 5 \ \mathrm{Hz}$. To identify the relevant swimmer properties of both regimes, we simulate the magnetocapillary system within a Lissajous-type magnetic field in the absence of remanent magnetic moments as well as without deforming forces (Fig. \ref{fig_mechanism_Lissajous}). As in the case of a rotating magnetic field, the low-frequency peak is well recovered in the absence of swimmer deformations, while the high-frequency peak is recovered in the case of absent remanent moments. Consequently, the high-frequency regime must result from the metachronal swimmer deformations. 

\begin{figure}
\includegraphics[scale=1]{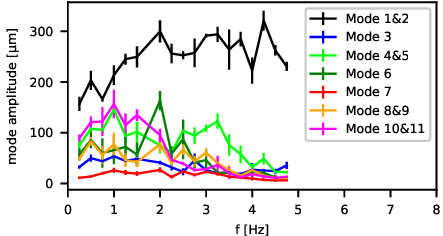}
\caption{Excitation of the eigenmodes of the magnetocapillary swimmer in the Lissajous field in the experiments, degenerate eigenmodes associated to the same eigenvalue are combined into a single curve. The lines are a guide for the eye.}
\label{fig_eigenmodes_Lissajous_exp}
\end{figure}
In contrast to the experiments with a rotating magnetic field, in the case of the Lissajous field more eigenmodes are excited (Fig. \ref{fig_eigenmodes_Lissajous_exp}). This is a result of the higher horizontal magnetic field amplitude, $B_h = 1.2 \ \mathrm{mT}$, used in these experiments. 
The three pairs of degenerate modes formed by modes 1 and 2, 4 and 5 as well as modes 10 and 11 are typically excited strongest. 

\begin{figure}
\includegraphics[scale=1]{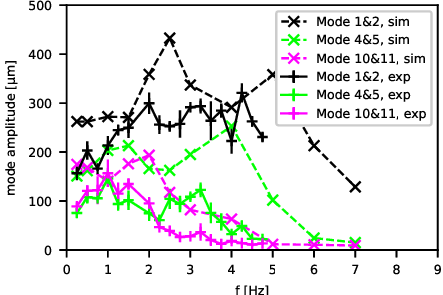}
\caption{Comparison of the excitation of the three main modes between experiments (pluses, solid lines) and simulations (crosses, dashed lines). The lines are a guide for the eye.}
\label{fig_eigenmodes_Lissajous_comp}
\end{figure}

Comparing the eigenmode excitation amplitudes between experiments and simulations, we observe a qualitative agreement in Fig. \ref{fig_eigenmodes_Lissajous_comp}. Both the order of magnitude and the trend of the mode amplitude with respect to the frequency are recovered. Quantitative differences between the experiments and the simulations may arise from the restricted time resolution of the experimental video tracking as well as from effects of the liquid interface. In particular, the interface leads to a reduced effective viscosity of the system, explaining why the experimental system generally shows higher excitation amplitudes than the simulated system assuming bulk viscosity.

\bibliography{supplementary}